# Minor Structural Differences Tune Functional Specialization of Antenna Subunits in the PSII Energy Transfer Network

*Johanna L. Hall*[a,b,c], *Erin NewRingeisen*[d], *Trisha Bhagde*[a,b], *Arnold M. Chan*[a], *Krishna K. Niyogi*[b,e,f,g], *Masakazu Iwai*[b,e], *Graham R. Fleming*[a,b,c,*]

[a]Department of Chemistry, University of California, Berkeley, Berkeley, CA 94720, USA.

[b]Molecular Biophysics and Integrated Bioimaging Division, Lawrence Berkeley National Laboratory, Berkeley, CA 94720, USA.

[c]Kavli Energy Nanoscience Institute at Berkeley, Berkeley, CA 94720, USA.

[d]Department of Molecular and Cell Biology, University of California, Berkeley, Berkeley, CA 94720, USA.

[e]Department of Plant and Microbial Biology, University of California, Berkeley, Berkeley, CA 94720, USA.

[f]Howard Hughes Medical Institute, University of California, Berkeley, Berkeley, CA 94720, USA.

[g]Innovative Genomics Institute, University of California, Berkeley, Berkeley, CA 94720, USA.

ABSTRACT

Photosystem II (PSII) contains a diverse array of pigment-protein subunits with specialized roles in the antenna network. We investigate the functional significance of this subunit diversity in two structurally similar yet functionally distinct PSII subunits: minor light-harvesting complex CP29 and major light-harvesting complex LHCII. We combine two-dimensional electronic-vibrational spectroscopy with lifetime density analysis and structure-based kinetic modeling to assign spectral features to exciton states, providing molecular detail to probed spectral features. We show how the distinct pigment compositions of CP29 and Lhcb1, a monomeric subunit of the major LHCII trimer, reshape excitonic connectivity and energy transfer pathways in the complexes. Compared to Lhcb1, energy transfer pathways in CP29 are more spatially connected and reversible, suggesting CP29 evolved to distribute excitation between the core and periphery of the PSII antenna. We find that small structural modifications from a common framework tune CP29 and Lhcb1 to complementary roles in the PSII supercomplex.

MAIN TEXT

Photosystem II (PSII) is the primary light-harvesting complex in oxygenic photosynthesis and is the only natural system capable of oxidizing water to evolve molecular oxygen[1]. It can harvest photons with near-unity quantum efficiency in ideal conditions and dissipate excess energy via non-photochemical quenching (NPQ) under light stress[1]. PSII is composed of distinct protein subunits[2,3] rather than multiple copies of structurally similar light-harvesting units, which differentiates its architecture and energetic landscape from other photosystems[4–6]. The reason for this unique design remains unclear, but is hypothesized to reflect the diverse functional requirements of PSII, including efficient energy trapping, water oxidation, and photoprotection[7]. Each PSII subunit likely has a specific role in energy transfer and trapping, NPQ, antenna reorganization, and even macro-organization, such as PSII array formation and thylakoid membrane stacking. However, the distinct functional roles of individual subunits are still under investigation[8–12].

To better understand these functional distinctions in the context of energy transfer, we investigate two types of PSII light-harvesting complexes that share many similarities in their tertiary structures while occupying distinct positions in the PSII supercomplex: CP29, a minor light-harvesting complex, and LHCII, the major light-harvesting complex (Figure 1). These subunits are both encoded by the LHC gene family[13,14] and display remarkably similar pigment arrangements and protein backbones, especially in the transmembrane core domain[2] (Figure 2A). Numerous studies characterizing CP29 have compared it with major LHCII due to this structural similarity[15–20]. However, a number of structural differences still distinguish these complexes, most notably the differences in the type of chlorophyll occupying conserved binding sites, as specified below. We aim to understand how these differences specialize the energy-transfer properties of these otherwise highly similar antenna proteins. To resolve these properties, we intend to compare the energy transfer dynamics of CP29, a monomeric LHCII, with Lhcb1, a monomeric subunit within the major LHCII trimer, at the individual chlorophyll level (Figure 1B). This requires a method to assign spectroscopic features to individual excitonic states and determine how those states evolve following excitation. Here, we perform two-dimensional electronic-vibrational (2DEV) spectroscopy on reconstituted CP29 and Lhcb1 samples. 2DEV correlates electronic excitation and vibrational detection frequencies, providing independent spectral dimensions that improve the resolution of congested states[21]. While the enhanced spectral resolution of 2DEV has been previously demonstrated[22–26], its specific ability to map excitonic states onto the spatial domain has not been fully utilized. We present a method that takes advantage of this mapping to assign 2DEV spectral peaks to specific chlorophyll states using the characteristic timescales in the signal and a quantum dynamical model. With this method, we can gain a greater level of molecular insight from our spectroscopic studies of these systems.

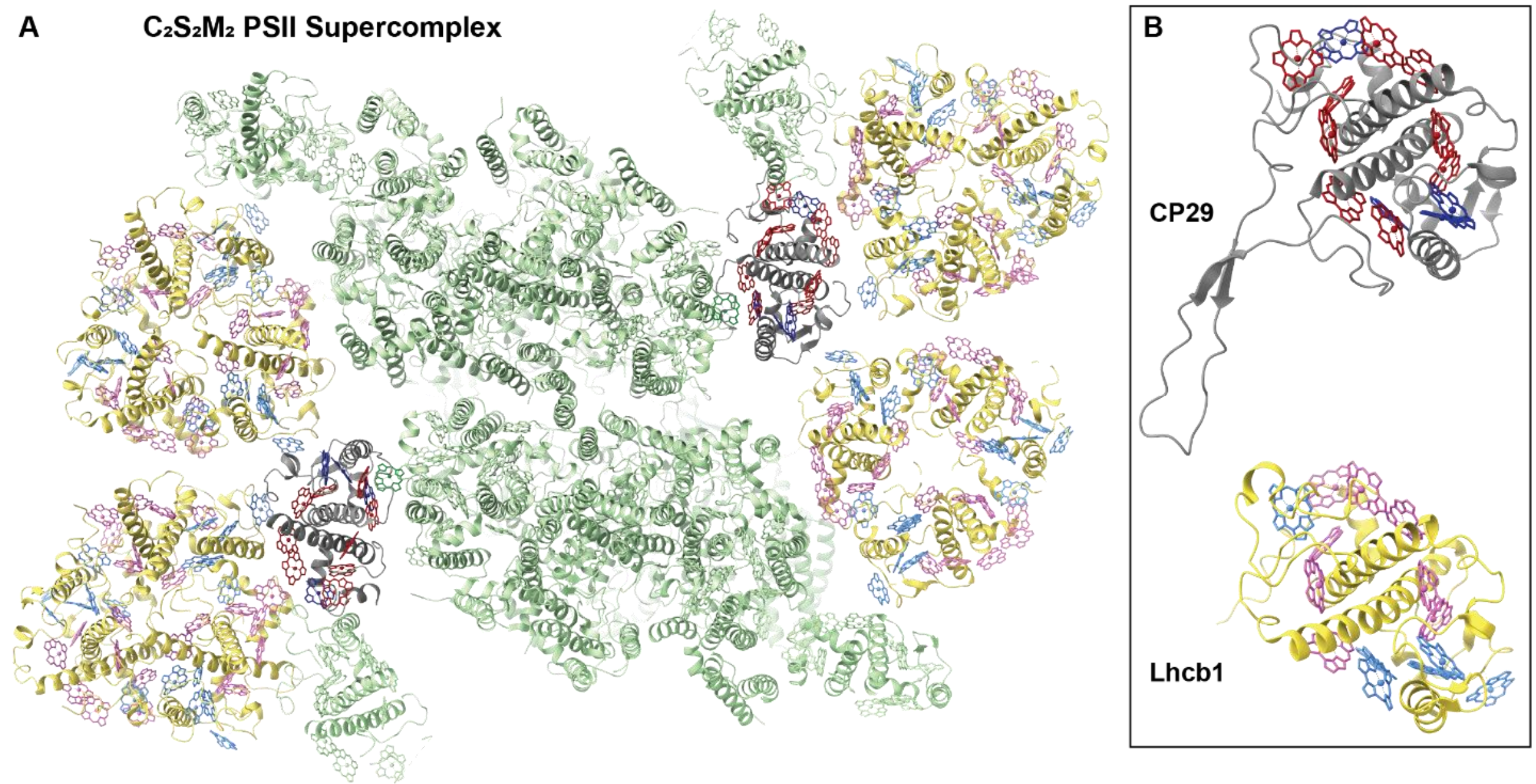

**Figure 1.** (A) The $C_2S_2M_2$-type PSII supercomplex (PDB 5XNL). The CP29 subunits are color-coded with grey protein backbones, dark red Chls *a*, and dark blue Chls *b*. The major LHCII subunits are color-coded with yellow protein backbones, pink Chls *a*, and light blue Chls *b*. All other subunits and chlorophylls are colored light green. (B) Enlarged CP29 and Lhcb1 monomeric protein structures to demonstrate structural similarity.

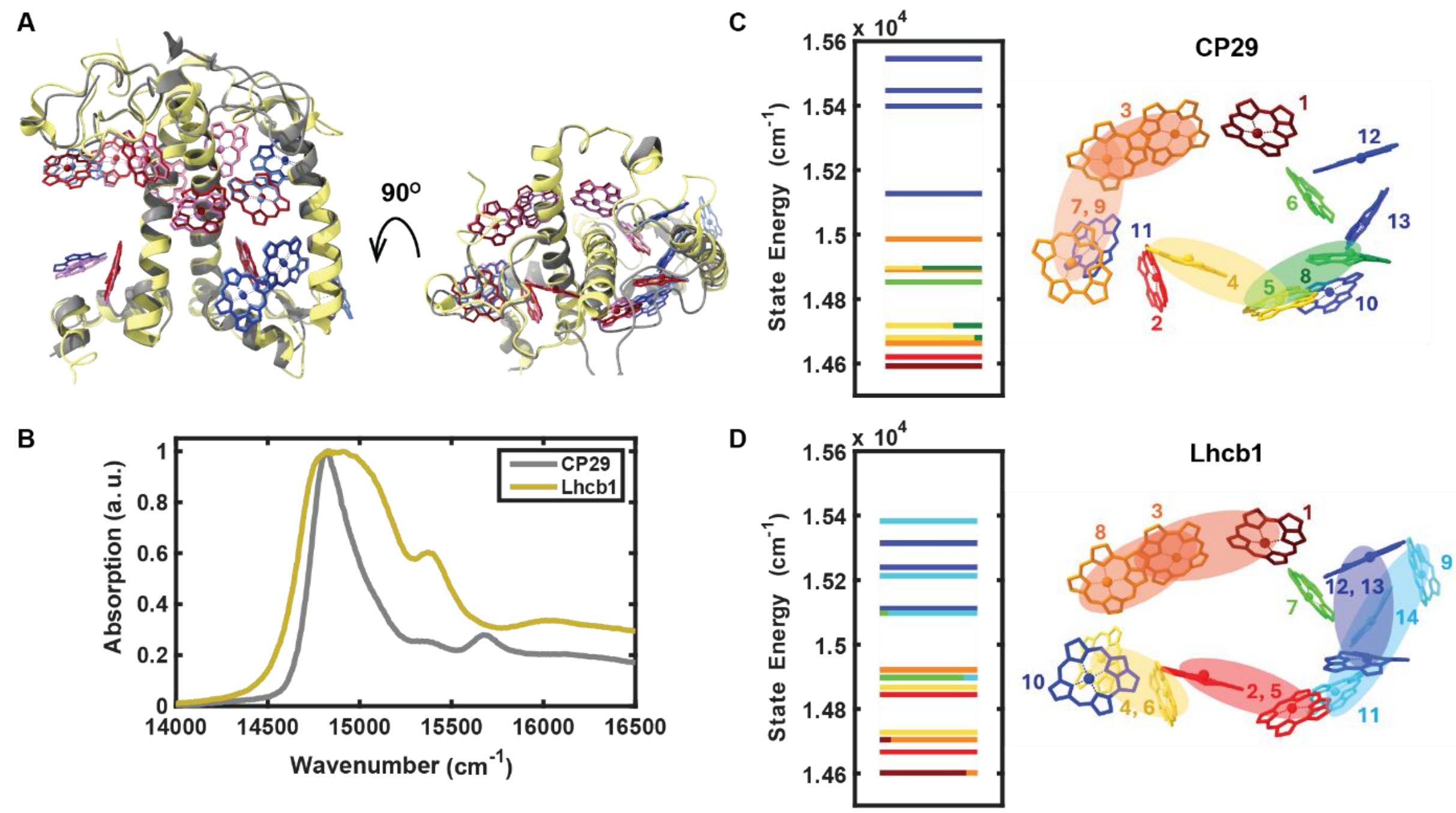

**Figure 2**. (A) Structural overlay of CP29 (grey protein backbone, dark blue Chls *b*, and dark red Chls *a*) and Lhcb1 (yellow protein backbone, light blue Chls *b*, and pink Chls *a*) in two orientations (PDB 5XNL). (B) Linear absorption spectra of CP29 (grey) and Lhcb1 (yellow) in $D_2O$ and deuterated glycerol at 85K, a.u. arbitrary units. (C-D) Calculated exciton energies and their projection on the structure for CP29 (C) and Lhcb1 (D), with exciton numbering starting at the lowest-energy state. Light- or dark-blue colored chlorophylls are chlorophyll *b*; all other-colored chlorophylls are chlorophyll *a*.

Although major LHCII has been extensively characterized[26–33], the excitonic structure and energy-transfer pathways of CP29 remain less well characterized. The excitonic ordering, energy-transfer directions, and functional role of CP29 within the PSII supercomplex remain debated[15,17,19,20,34–43]. The structural differences between CP29 and Lhcb1 largely stem from the identity of bound pigments at conserved sites, which likely arise from hydrogen bonding between chlorophyll and binding-site residues[16]. Lhcb1 binds 14 chlorophylls (Chls), including eight Chl *a* and six Chl *b* molecules, whereas CP29 contains 13 Chls, including nine Chl *a* and four Chl *b* molecules[2,18,44]. The pigment occupying the 614 site is Chl *a* in Lhcb1 and Chl *b* in CP29, while the 601 and 609 sites contain Chl *b* in Lhcb1 and Chl *a* in CP29 (Figure 3E). Lhcb1 also binds Chl *b* 605, which has no corresponding pigment in CP29. We note that the 5XNL structures used here reflect the most probable pigment composition and location in CP29 and Lhcb1. Still, conformational and pigment-binding variation may exist across different structures. Additional structural differences between CP29 and Lhcb1 are discussed in the SI. In this work, we employ a combination of 2DEV spectroscopy, quantum dynamical simulation, and lifetime density analysis to reveal how small structural changes can be utilized to produce differing and complementary functionality in CP29 and Lhcb1.

The linear absorption spectra for CP29 and Lhcb1 at 85K are shown in Figure 2B. The Chl *b* peak in CP29 is significantly blue-shifted and less intense compared to the corresponding peak in Lhcb1[36]. We also observe significant inhomogeneous broadening in the Lhcb1 spectrum compared to the trimeric major LHCII[26,30], which we attribute to increased structural heterogeneity associated with removal from the native trimer. This reduces excited-state resolution along the excitation axis of the 2DEV spectra, but the vibrational peaks are still well resolved (Figure 3B). Analysis of the excited-state absorption (ESA) and ground-state bleach (GSB) features along the detection axis reveals strong similarities between the CP29 and Lhcb1 spectra, consistent with their highly similar pigment organizations. The energetic separation of the Chl *a* and Chl *b* $Q_y$ transitions enables detection-axis slices at selected excitation frequencies to distinguish vibrational features arising predominantly from Chl *a* or Chl *b* states (Figure S1).

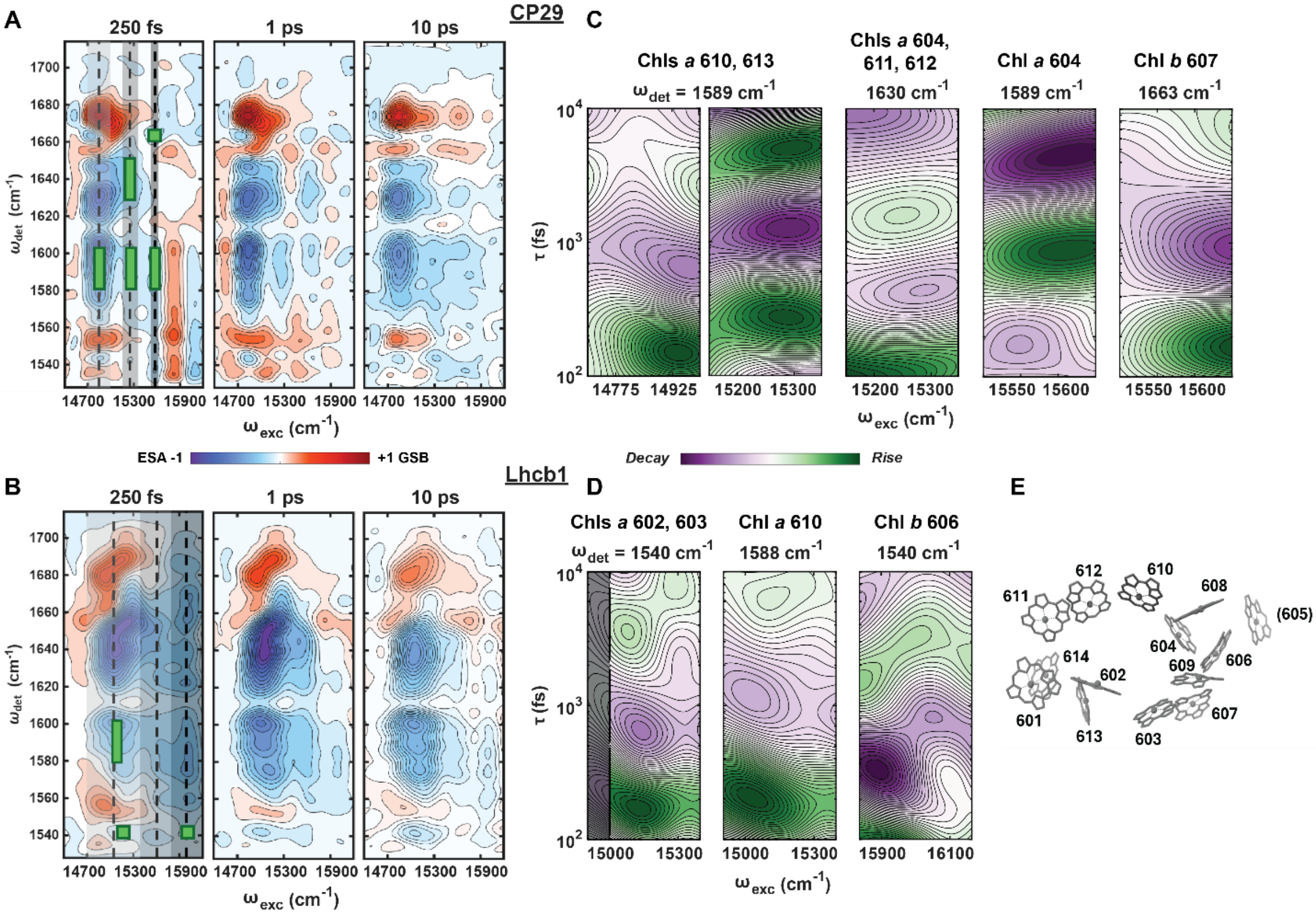

**Figure 3**. 2DEV data and peaks assigned based on LDA. (A-B) 2DEV data for CP29 (A) and Lhcb1 (B). Grey regions on the T = 250 fs waiting time map indicate the excitation frequency regions investigated for each detection peak, with higher-energy regions corresponding to darker shading. Dashed lines denote the center frequency of these regions. Green boxes correspond to the peaks assigned using the lifetime density maps (LDMs) shown in (C) for CP29 and (D) for Lhcb1. Chlorophyll labels for both complexes are shown in (E).

To further assign individual vibrational features to specific chlorophylls, we analyzed the temporal evolution of the 2DEV signals using lifetime density analysis (LDA), which is particularly suited for kinetic descriptions of biologically and spectrally complex systems like photosystems[45]. LDA represents the data as a semi-continuous sum of lifetimes weighted by their contribution to the overall signal dynamics, without requiring a predetermined number of kinetic components. Because LDA was applied independently at each spectral feature, the resulting lifetime distributions could vary across the spectral range, allowing complex rise and decay dynamics to be captured without constraining the entire spectrum to a common set of time constants, as in conventional global target analysis. We applied LDA to the time-dependent signal at each vibrational feature using 200 logarithmically spaced lifetimes between 10 fs and 10 ps, after which the LDA became numerically unstable because the faster energy-transfer dynamics were largely complete and the remaining signal was dominated by slower population decay. The resulting lifetime densities were multiplied by the sign of the corresponding signal (-1 for ESA and +1 for GSB) to distinguish characteristic timescales associated with population rise and decay[45]. Only ESA features were analyzed for these assignments since they directly report on excited-state dynamics, whereas GSB features are convoluted with information about other pathways including vibrational relaxation and energy transfer to states different from the chlorophyll S1 state of interest. The lifetime densities were combined over specified excitation frequency ranges to generate lifetime density maps (LDMs).

To aid in assigning vibrational features to individual chlorophylls, we also modeled the energy transfer dynamics using structure-based kinetic models of CP29 and Lhcb1. These models were constructed following previous descriptions[23,46,47] using Hamiltonians from Novoderezhkin and coworkers[36,48]. The exciton states obtained from these calculations and their projection onto the site basis for CP29 and Lhcb1 are shown in Figure 2C and D[49]. We used the computed rate matrices for each subunit to simulate population dynamics resulting from a Gaussian excitation ($\mu = 0$ fs, $\sigma = 100$ fs) at each experimentally measured excitation frequency. These dynamics were projected onto the site basis and analyzed using LDA to obtain the characteristic kinetic components associated with each chlorophyll. Simulated LDMs for each chlorophyll were compared with experimentally derived LDMs of vibrational peaks in the corresponding excitation-frequency ranges. Experimental vibrational features were assigned to specific chlorophylls when their LDMs exhibited rise and decay components whose timescales matched those of the simulated chlorophyll populations (Figure S2). The excitation frequency windows analyzed and resulting peak assignments are summarized in Figure 3A and B and Table 1. Using this approach, five vibrational features were assigned in CP29 and three in Lhcb1 (Figure 3, Table 1). The inhomogeneous broadening in the Lhcb1 spectrum complicated further peak assignments. In both complexes, most assigned peaks correspond to the lower-energy excitonic states, which is expected because population accumulates in these states over the course of the dynamics, making their vibrational signatures the most readily detected.

In CP29, the lowest-energy excitonic states are primarily formed by Chl *a* 610 and Chl *a* 613, which are both assigned to the features around ($\omega_{exc}$ = 14850 $cm^{-1}$, $\omega_{det}$ = 1589 $cm^{-1}$) and ($\omega_{exc}$ = 15250 $cm^{-1}$, $\omega_{det}$ = 1589 $cm^{-1}$). The next-lowest energy excitonic state is primarily formed by Chls *a* 611 and 612, which are assigned to the feature around ($\omega_{exc}$ = 15250 $cm^{-1}$, $\omega_{det}$ = 1630 $cm^{-1}$). Higher-energy Chl *a* 604 was also assigned to this feature. In Lhcb1, the lowest-energy excitonic state is primarily formed by Chl *a* 610, assigned to the feature around ($\omega_{exc}$ = 15100 $cm^{-1}$, $\omega_{det}$ = 1588 $cm^{-1}$). The next-lowest energy excitonic state is primarily formed by Chls *a* 602 and 603, which were assigned to the feature at ($\omega_{exc}$ = 15100 $cm^{-1}$, $\omega_{det}$ = 1540 $cm^{-1}$). Some spectral features at higher-energy excitation frequencies were further assigned in both subunits. In CP29, Chl *a* 604 was found to dominate the feature around ($\omega_{exc}$ = 15600 $cm^{-1}$, $\omega_{det}$ = 1589 $cm^{-1}$) and Chl *b* 607 was found to dominate the feature at ($\omega_{exc}$ = 15600 $cm^{-1}$, $\omega_{det}$ = 1663 $cm^{-1}$). In Lhcb1, Chl *b* 606 was found to dominate the feature at ($\omega_{exc}$ = 16000 $cm^{-1}$, $\omega_{det}$ = 1540 cm-1 $cm^{-1}$).

**Table 1**. Summary of chlorophyll assignments to 2DEV spectral features

| Subunit | Center $\omega_{exc}$ of spectral feature [cm$^{-1}$] | Center $\omega_{det}$ of spectral feature [cm$^{-1}$] | Chlorophyll assignment |
|---|---|---|---|
| **CP29** | 14850 | 1589 | Chls *a* 610, 613 |
| | 15250 | 1589 | Chls *a* 610, 613 |
| | 15250 | 1630 | Chls *a* 604, 611, 612 |
| | 15600 | 1589 | Chl *a* 604 |
| | 15600 | 1663 | Chl *b* 607 |
| **Lhcb1** | 15100 | 1588 | Chl *a* 610 |
| | 15100 | 1540 | Chls *a* 602, 603 |
| | 16000 | 1540 | Chl *b* 606 |

Peaks that exhibit complementary rise and decay kinetics can reveal energy transfer pathways[23]. In CP29, the peak assigned to Chl *b* 607 exhibits a decay between approximately 500 fs and 2 ps, while the peak assigned to Chl *a* 604 exhibits a rise over the same timescale with a similar excitation-frequency dependence (Figure 3A). This suggests energy transfer from Chl *b* 607 to Chl *a* 604 between 500 fs and 2 ps in CP29. We also observe that the peak assigned to Chls *a* 610 and 613 in the second excitation-frequency window exhibits a decay between 0.9 and 2 ps, while the peak assigned to Chls *a* 604, 611, and 612 exhibits a rise on the same timescale and excitation-frequency dependence. This suggests energy transfers from Chls *a* 610 and 613 to Chls *a* 604, 611, and 612 between 0.9 and 2 ps.

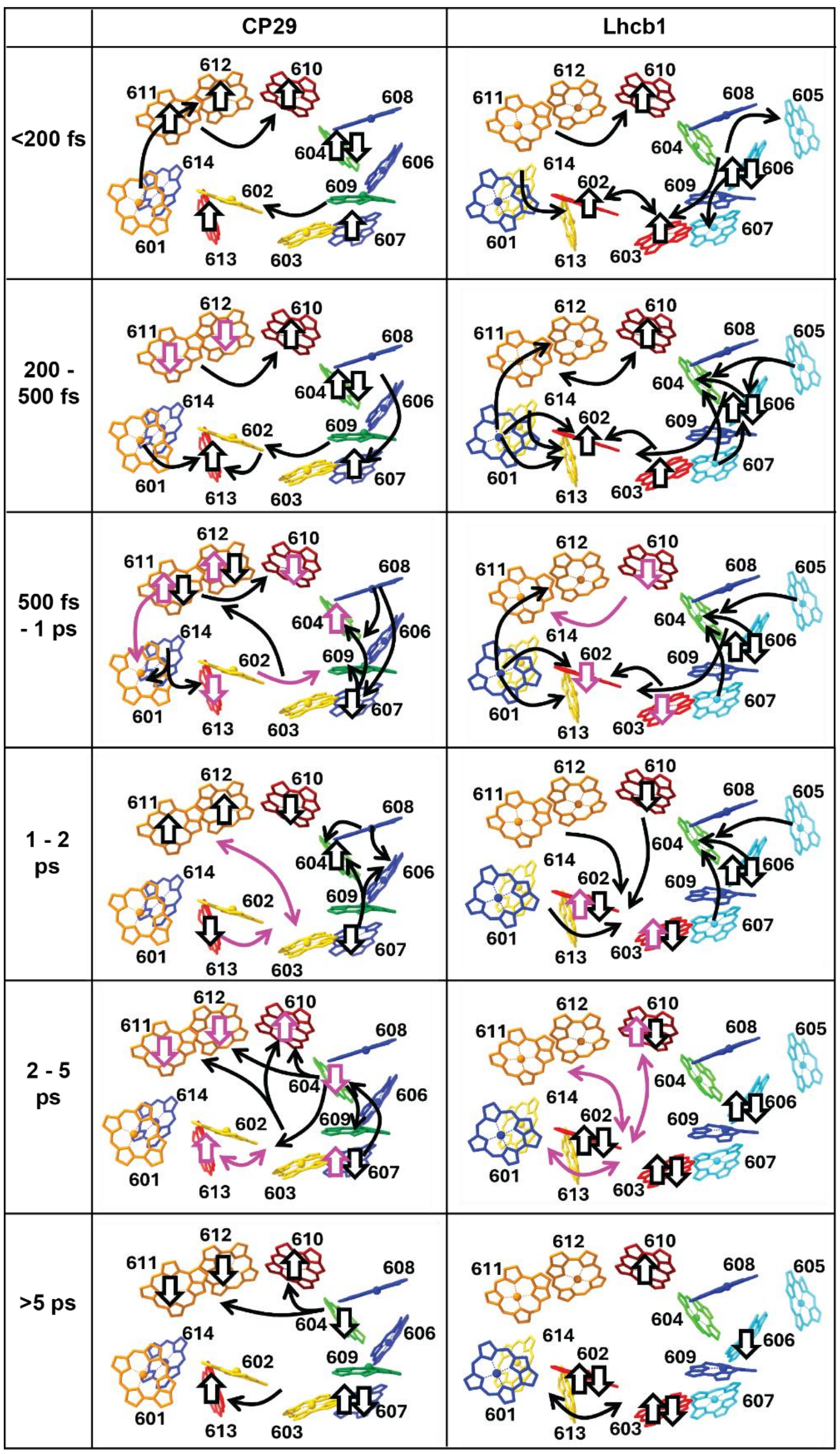

**Figure 4**. Summary of energy transfer dynamics in CP29 and Lhcb1. Up and down arrows (⇑ or ⇓) indicate rises and decays, respectively, from assigned experimental peaks. Curved arrows indicate energy transfer pathways revealed by simulation. Pink coloration of any arrow indicates that the arrow has changed direction since the previous time period. Light- or dark-blue colored chlorophylls are chlorophyll *b,* and any other-colored chlorophylls are chlorophyll *a*.

The simulated LDMs were analyzed using the same approach to identify the energy-transfer pathways predicted by the kinetic model. Figure 4 summarizes the experimentally inferred population rises and decays (vertical arrows) and the simulation-predicted energy transfer pathways (curved arrows) for CP29 and Lhcb1. Note that the energy transfer pathways predicted by the kinetic model reflect the directions of net energy transfer irrespective of path. Pink arrows denote rises, decays, or energy transfer pathways that reverse direction relative to the preceding time window. Overall, both the experimental and simulated dynamics indicate that excitation energy transfer events are distributed across multiple pathways and timescales[47]. LDA further demonstrates that transfer between the same pair of pigments can occur on multiple characteristic timescales, highlighting the dynamical complexity that may be obscured by conventional global analysis. We also observe frequent reversals in population flow between consecutive time windows. This can arise from exciton delocalization or close energetic spacing within thermal energy between participating states. Even for close energetic spacings, the forward and reverse energy transfer rates between two given states differ by a Boltzmann factor, resulting in a net timescale difference between the forward and reverse processes. This, coupled with the multitude of energy transfer options available to each excited state, results in energy transfer dynamics that are surprisingly complex in both the time and spatial domains.

CP29 and Lhcb1 share several common energy transfer pathways, including transfer from Chls *b* to Chl *a* 604, followed by transfer to lower-energy Chl *a* states. This role of Chl *a* 604 as an intermediate in energy transfer has been previously reported[16,36,50–52]. Despite these similarities, the two complexes exhibit important differences in their energy transfer dynamics. These differences are particularly evident in the interactions between Chl 609 and Chls *a* 602/603, and between Chl 601 and Chls *a* 611/612 (Figure 4). Within the first 500 fs, both complexes exhibit net transfer from Chl 609 to Chls *a* 602/603 and from Chl 601 to Chls *a* 611/612. These transfer directions persist between 500 fs and 1 ps in Lhcb1. In contrast, CP29 exhibits a reversal of net population flow over the same interval, with excitation transferring from Chls *a* 602/603 back to Chl 609 and from Chls *a* 611/612 back to Chl 601. These reversals are made possible through the distinct pigment composition of CP29 compared to Lhcb1. Because Chls *a* occupy the 609 and 601 binding sites in CP29, which are instead Chls *b* in Lhcb1, exciton delocalization and bidirectional energy transfer between these chlorophylls and other Chls *a* are more energetically feasible in CP29 than in Lhcb1. We discuss the functional implementation of these differences below.

In this work, we demonstrate how the spatial mapping provided by 2DEV spectroscopy can be combined with lifetime density analysis and structure-based kinetic modeling to assign spectral features to exciton states in spectrally congested systems. Because the lower-energy exciton states are the most readily detected, particularly in lower-energy excitation regions, excitation-frequency resolution is necessary to resolve higher-energy exciton states. These assignments also rely on spectral resolution along an axis encoding spatial information, provided by the mid-IR detection axis probing the carbonyl stretch of chlorophyll molecules in this experiment.

This approach provides a means of connecting spectral features to specific excitonic states through their population dynamics, providing molecular-level detail to observed signals. This insight can be useful for many applications. It can distinguish energetically overlapping excitonic states when they exhibit different spatial distributions or temporal dynamics, identify less prominent spectral features through their characteristic temporal evolution, and characterize the states probed by the spectral signatures of specific complexes within larger systems.

Applying this approach to CP29 and Lhcb1 allowed us to assign and observe the population dynamics of six chlorophylls in CP29 and four in Lhcb1 despite the spectral congestion in these complexes. Chl *a* 610 was assigned to the 1576–1602 $cm^{-1}$ feature in the lowest-energy excitation-frequency window in both complexes, indicating that this feature reflects Chl *a* 610 dynamics despite differences in pigment composition and structure. This provides a potential spectroscopic reporter of the lowest-energy exciton in CP29 and Lhcb1 in larger systems. From this analysis, we predict the vibrational feature around ($\omega_{exc}$ = 14800 $cm^{-1}$, $\omega_{det}$ = 1576-1602 $cm^{-1}$) in the major LHCII trimer to also report on the dynamics of its lower-energy excitons, including Chl *a* 610.

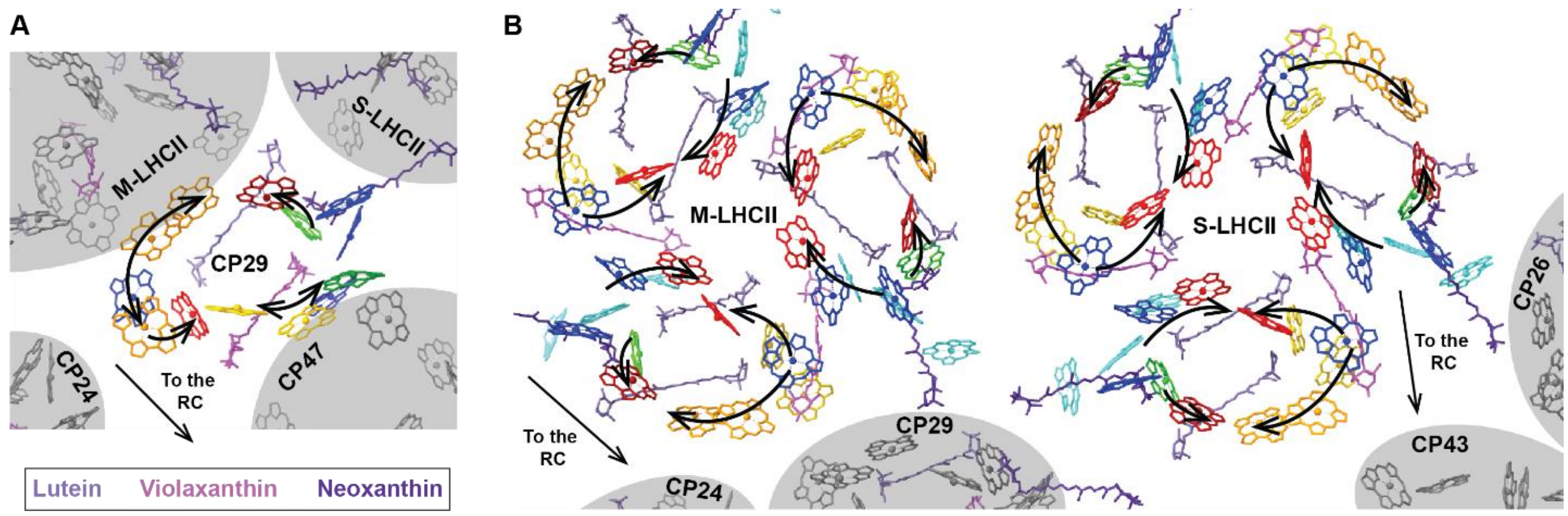

**Figure 5**. CP29 (A) and major LHCII trimers (B) in the $C_2S_2M_2$ PSII structure with lutein, violaxanthin, and neoxanthin carotenoids shown (PDB 5XNL). The dominant energy transfer pathways in either complex are depicted with unidirectional or bidirectional arrows.

Both experiments and structure-based kinetic modeling revealed energy transfer dynamics in CP29 and Lhcb1 occurring across multiple pathways and timescales. We also observe reversals in the net direction of population flow for several pathways. Together, these observations reveal a highly interconnected energy landscape, even at the subunit level. This landscape enables excitations to explore broader regions of the antenna, where photoprotective processes such as NPQ can dissipate energy when activated. It also allows excitations to reach other reaction centers (RCs), thereby allowing PSII to retain energy transfer and trapping ability in fluctuating environments.

While both complexes contained reversible energy transfer events, CP29 was observed to have more. These differences are associated with the identity of the chlorophylls at the 601 and 609 sites, which are Chls *a* in CP29 and Chls *b* in Lhcb1. This ultimately alters the excitonic landscape in CP29, producing states delocalized over Chls *a* 602/603/609 and Chls *a* 601/611/612. CP29 also contains two closely spaced lowest-energy excitons dominated by Chls *a* 610 and 613 (Figure 2C), consistent with previous reports[15,38]. These features create pathways through which excitation can redistribute from the central, lowest-energy states toward pigments on opposite sides of CP29 that interface with CP47 and M-LHCII (Figure 5A). In contrast, energy transfer in Lhcb1 is more strongly directed toward its lowest-energy pigment clusters, with limited redistribution toward peripheral chlorophylls (Figure 5B). This highlights the increased spatial connectivity of energy transfer pathways in CP29 compared to Lhcb1.

This increased connectivity provides a mechanistic basis for the behavior of CP29 within the larger PSII supercomplex. Previous simulations of the $C_2S_2M_2$ PSII supercomplex showed that an excitation originating in CP29 has approximately equal probabilities of being trapped by the RC within the same monomer of PSII and by the RC of the other monomer[53]. This implies that, on average, CP29 transfers the same amount of energy toward the core (through CP47) as it does toward the periphery (through M-LHCII). The reversible pathways identified here demonstrate how CP29 kinetically links its core-facing and peripheral regions.

Clearly, relatively small changes in chlorophyll composition alter the energy landscape and dynamics of population flow within the PSII antenna. Together, the increased connectivity, more prevalent reversible population flow, and balanced trapping probabilities in either RC suggest CP29 serves an important role in distributing energy between the core and antenna. Because CP29 connects multiple PSII subunits and serves as an intermediate step in many energy transfer pathways[47], it may be a good candidate for NPQ processes to catch and dissipate excess energy, as previously proposed[37].

These results demonstrate that functional specialization of PSII antenna proteins can be achieved through relatively minor structural changes. Rather than requiring large-scale structural rearrangements, modifying the chlorophyll identity at conserved binding sites can reshape excitonic delocalization, alter the directionality and reversibility of energy transfer, and adjust how an excitation explores the larger antenna network. CP29 and Lhcb1 appear to share a common structural framework but are tuned to complementary roles within PSII: Lhcb1 more strongly funnels excitation toward its lowest-energy states, whereas CP29 provides greater connectivity between core-facing and peripheral regions.

This tuning also highlights the importance of state connectivity as a design principle. CP29 serves as an interesting example of how small structural differences from Lhcb1 can shift the design from being relatively more enthalpy-driven to more entropy-driven with increased connectivity and reversibility[53]. Such tuning may represent a general strategy by which photosynthetic antenna proteins acquire distinct roles within complex light-harvesting assemblies. Working together, these complementary subunits allow PSII to accomplish its diverse goals of efficient energy trapping, water oxidation, and photoprotection.

ASSOCIATED CONTENT

pyLDM codes used for LDA are publicly available at Ref. 45.

AUTHOR INFORMATION

**Corresponding Author**

Graham R. Fleming

Hildebrand Hall 221, University of California, Berkeley, Berkeley, CA 94720

(510) 520-4220

**Author Contributions**

J.L.H conceived of the project, performed all experiments and simulations, analyzed and interpreted all data, and wrote the paper. G.R.F advised on the analysis of computational and experimental data, provided support for this work, and wrote the paper. E.N., A.M.C., and M.I. prepared the reconstituted CP29 and Lhcb1 samples. T.B. assisted with the Lhcb1 2DEV data collection. K.K.N. provided support for sample preparation. All authors reviewed the final manuscript.

**Funding Sources**

This research was supported by the US Department of Energy, Office of Science, Basic Energy Sciences, Chemical Sciences, Geosciences, and Biosciences Division. J.L.H. gratefully acknowledges support from the Kavli Energy Nanoscience Institute as a graduate fellow.

ACKNOWLEDGMENT

Molecular graphics and analyses were performed with UCSF ChimeraX, developed by the Resource for Biocomputing, Visualization, and Informatics at the University of California, San

Francisco, with support from National Institutes of Health R01-GM129325 and the Office of Cyber Infrastructure and Computational Biology, National Institute of Allergy and Infectious Diseases.

We thank Dr. Shiun-Jr Yang for helpful discussion.

# Supplemental Information for: Minor Structural Differences Tune Functional Specialization of Antenna Subunits in the PSII Energy Transfer Network

Johanna L. Hall[a,b,c], Erin NewRingeisen[d], Trisha Bhagde[a,b], Arnold M. Chan[a], Krishna K. Niyogi[b,e,f,g], Masakazu Iwai[b,e], Graham R. Fleming[a,b,c,*]

[a]Department of Chemistry, University of California, Berkeley, Berkeley, CA 94720, USA.

[b]Molecular Biophysics and Integrated Bioimaging Division, Lawrence Berkeley National Laboratory, Berkeley, CA 94720, USA.

[c]Kavli Energy Nanoscience Institute at Berkeley, Berkeley, CA 94720, USA.

[d]Department of Molecular and Cell Biology, University of California, Berkeley, CA 94720, USA.

[e]Department of Plant and Microbial Biology, University of California, Berkeley, CA 94720, USA.

[f]Howard Hughes Medical Institute, University of California, Berkeley, CA 94720, USA.

[g]Innovative Genomics Institute, University of California, Berkeley, CA 94720, USA.

**Table of Contents**

## Supplemental Figures

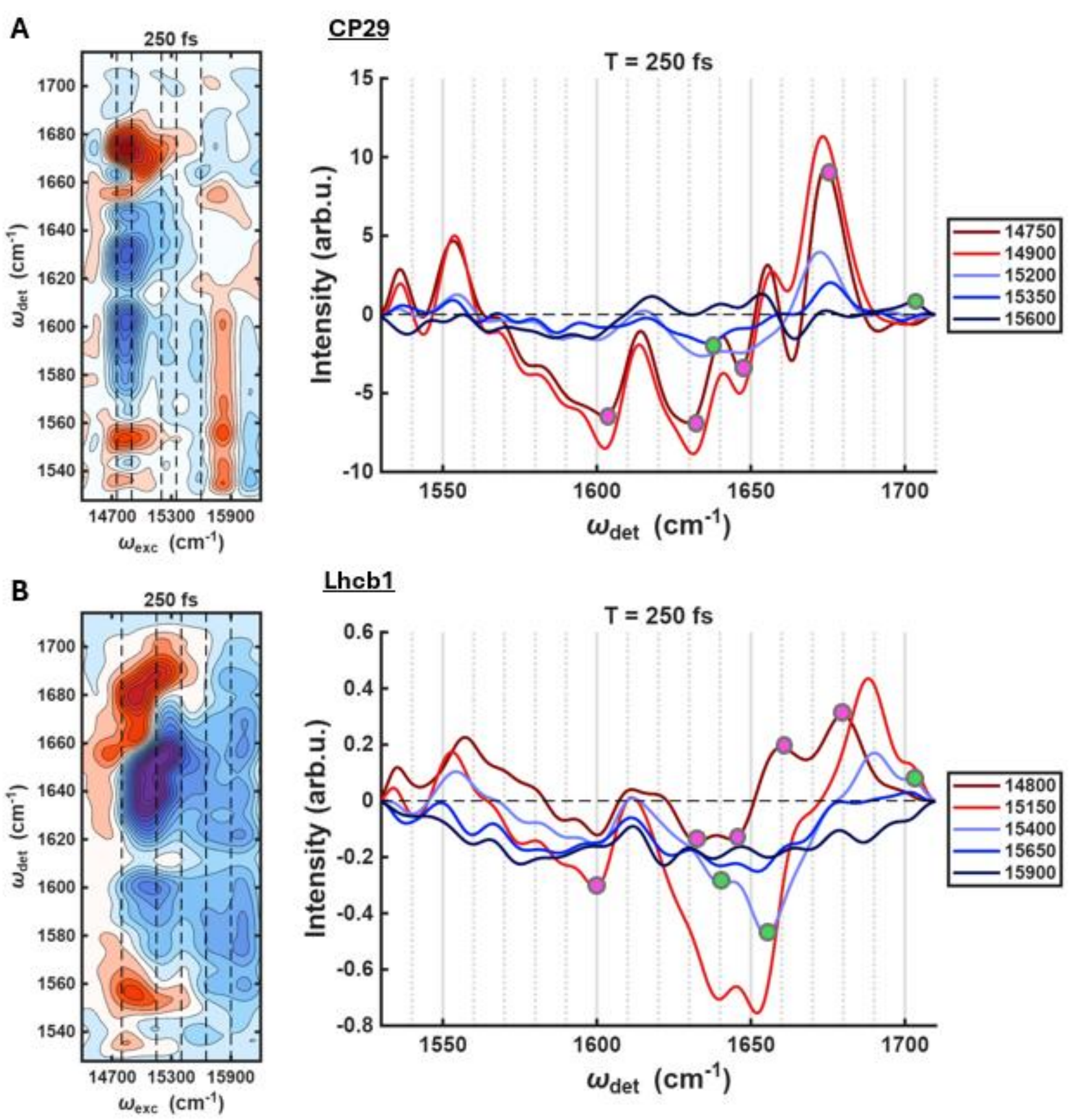

Figure S1. CP29 (A) and Lhcb1 (B) 2DEV spectra at T = 250 fs waiting time, with IR slices taken at the denoted excitation frequencies. Peaks labeled with pink dots are likely dominated by chlorophyll a, while peaks with green dots are likely dominated by chlorophyll b.

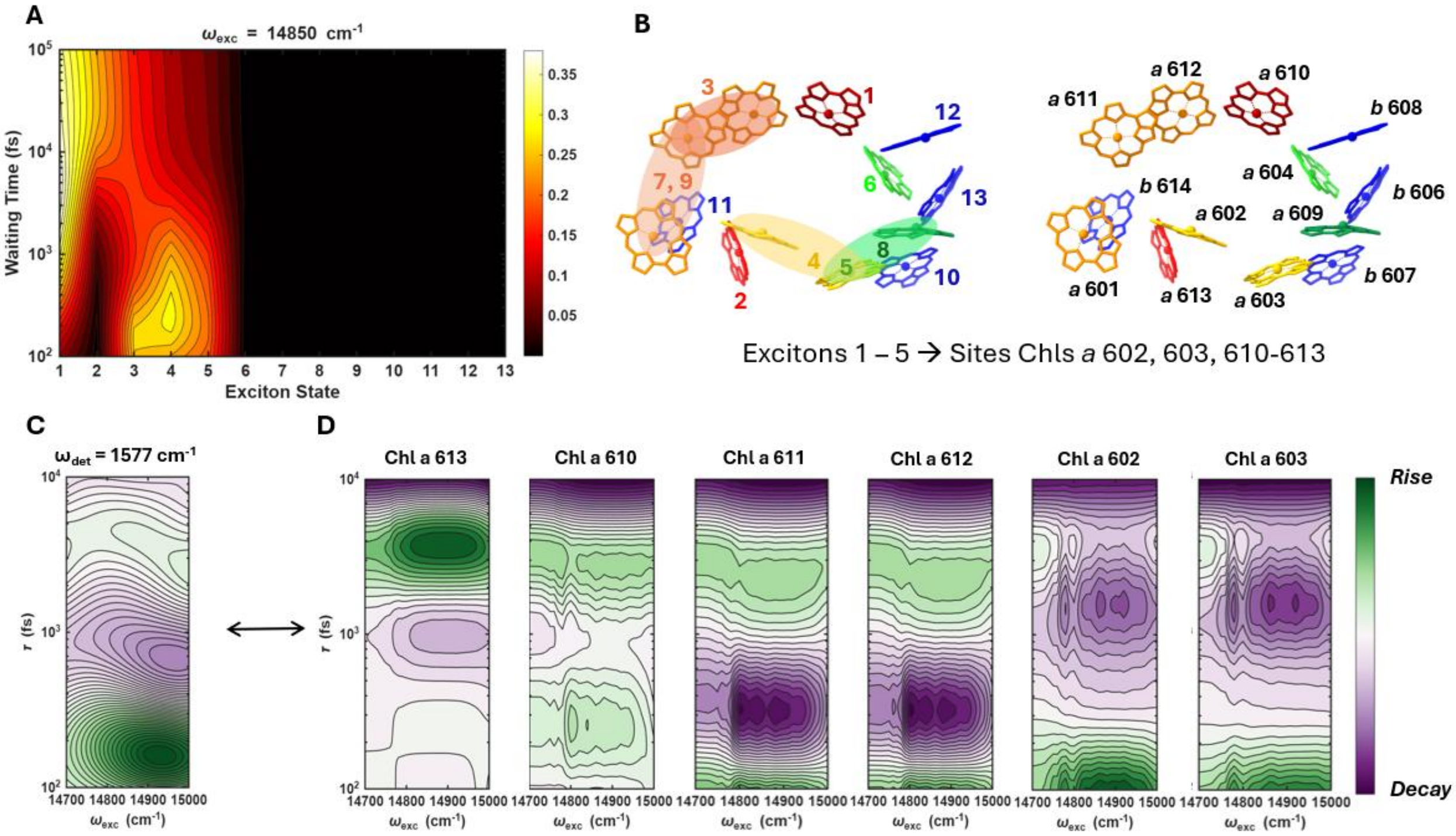

Figure S2. Experimental peak assignment procedure for the 1577 cm$^{-1}$ peak in CP29. (A) Simulated population evolution among excitons following a 14,850 cm$^{-1}$ excitation. (B) Exciton basis (left) and site basis (right). (C) LDM of experimental peak at $\omega_{det}$ = 1589 cm$^{-1}$ in the excitation frequency range centered at 14,850 cm$^{-1}$. (D) Simulated LDMs of the chlorophylls predicted to participate in the population evolution over the course of the measured dynamics.

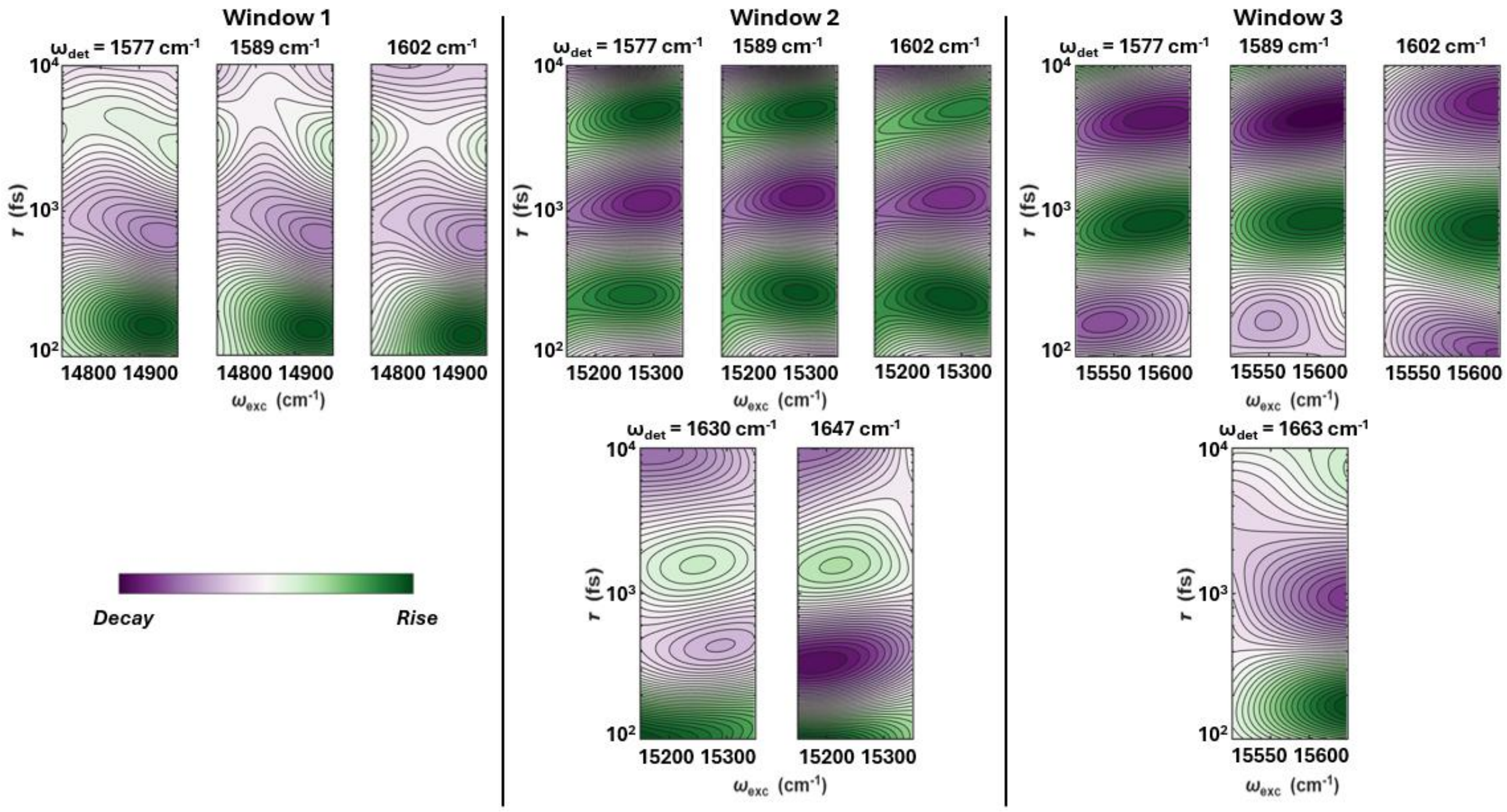

Figure S3. All CP29 experimental peaks assigned to chlorophyll states through LDA timescale comparisons.

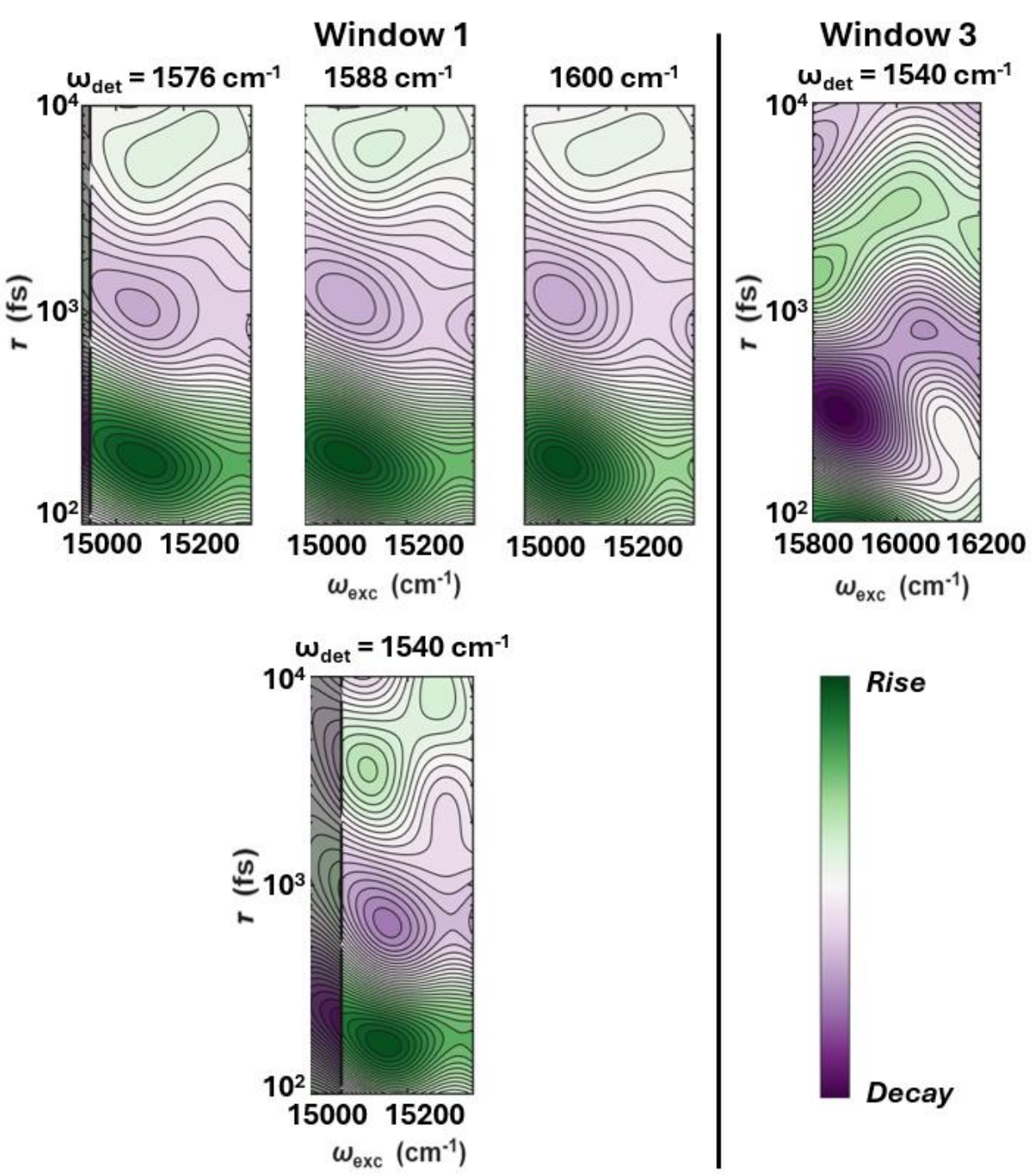

Figure S4. All Lhcb1 experimental peaks assigned to chlorophyll states through LDA timescale comparisons.

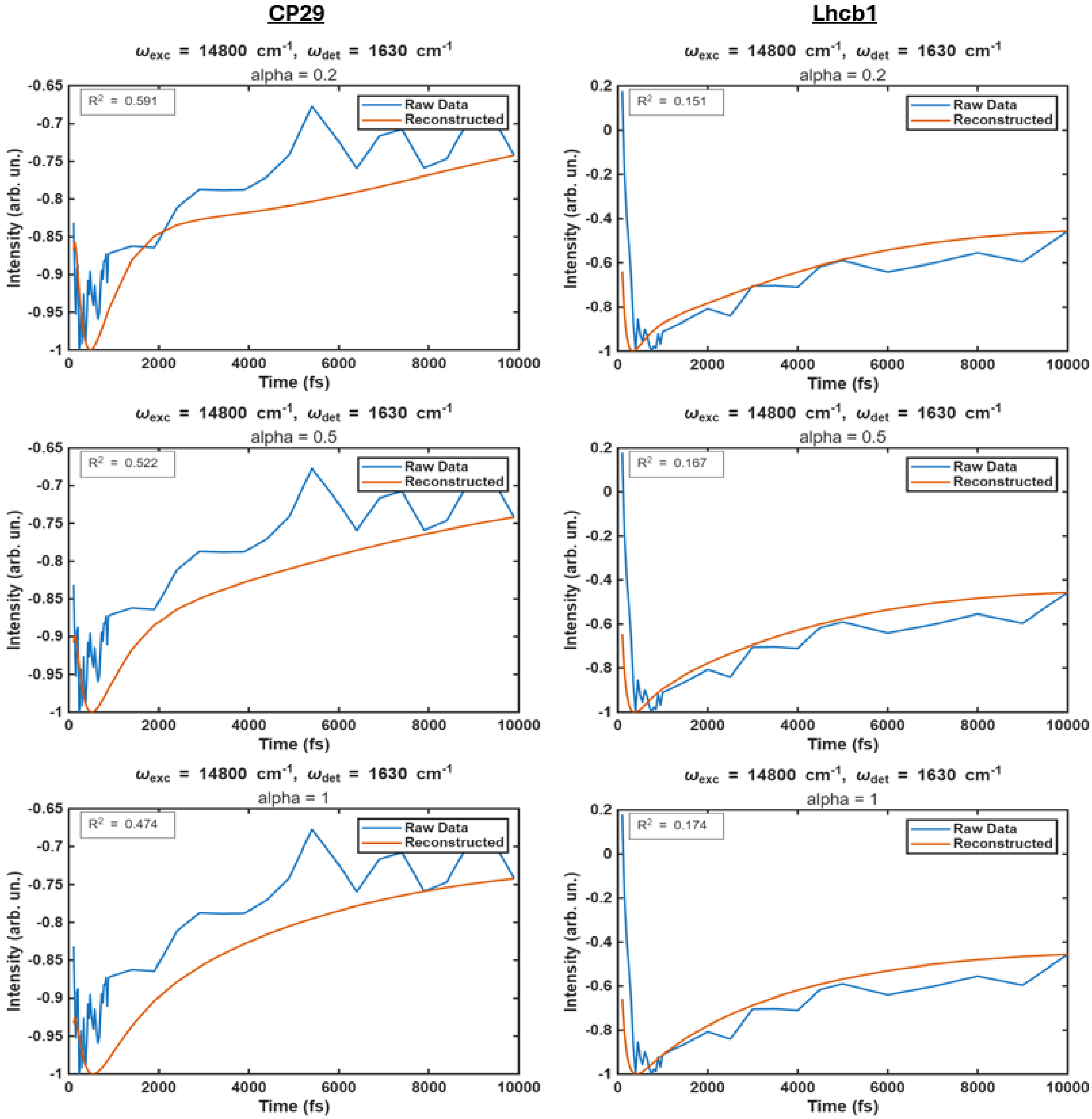

Figure S5. Time traces of CP29 2DEV signal (blue) at $\omega_{exc}$ = 14,800 cm$^{-1}$, $\omega_{det}$ = 1630 cm$^{-1}$ and $\omega_{exc}$ = 14,850 cm$^{-1}$, $\omega_{det}$ = 1602 cm$^{-1}$ and the time trace of the signal reconstructed from LDA parameters (orange) at three different alpha values.

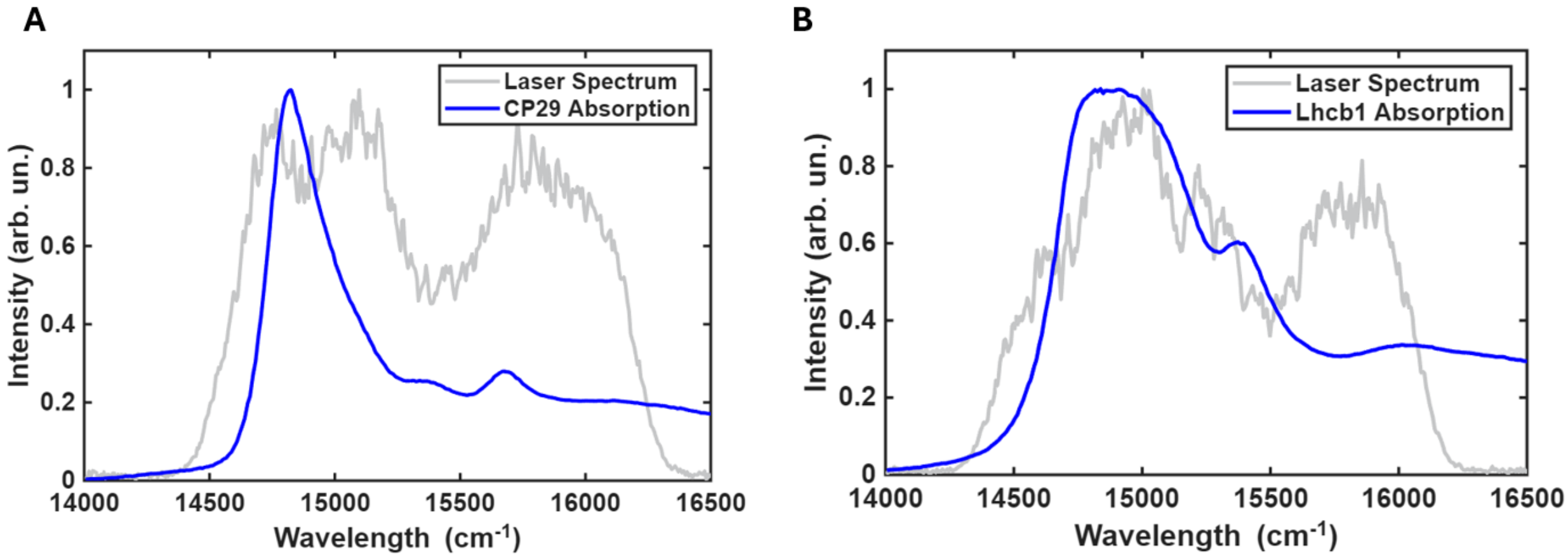

Figure S6. Laser excitation spectra (grey) and sample absorption spectra (blue) for CP29 (A) and Lhcb1 (B).

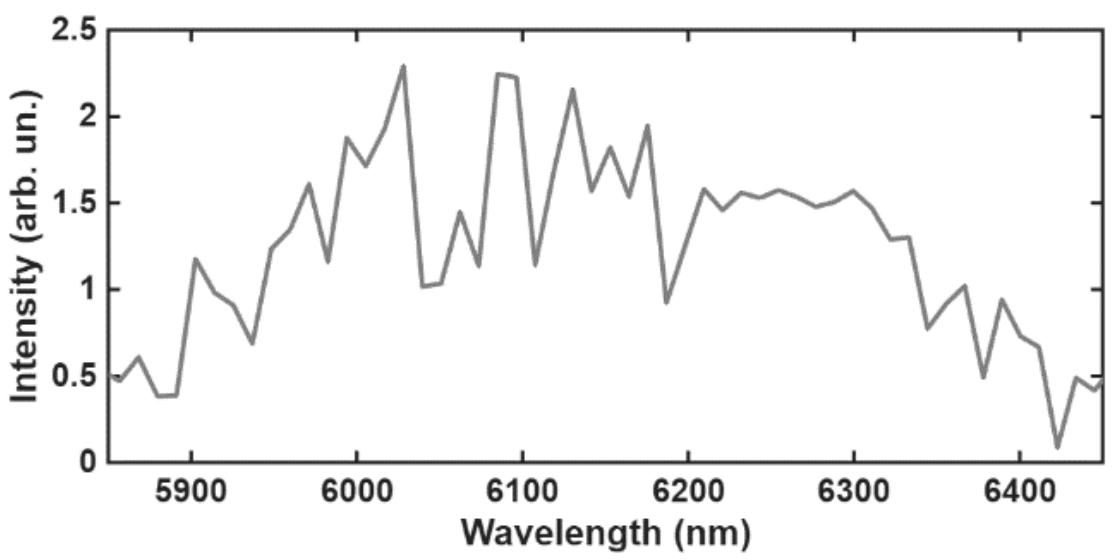

Figure S7. Mid-IR probe spectrum centered at 6100 nm.

## Structural Differences in CP29 and Lhcb1

Some structural studies of CP29 additionally report a Chl *a* 616 near the N-terminal region[19,20], which has been proposed to contribute to CP29-based NPQ[21]. However, other authors note that Chl *a* 616 is easily lost during protein purification[22] and that the deletion of Chl *a* 616 yields virtually no change in the fluorescence lifetime of the complex[23]. For these reasons, this pigment is not included in our reference CP29 Hamiltonian[11] and is therefore omitted in the present analysis.

Differences are also observed in carotenoid composition. One subunit of major LHCII binds four carotenoids, including two luteins (Lut) occupying the L1 and L2 sites, one neoxanthin (Neo) occupying the N1 site, and one violaxanthin (Vio) occupying the V1 site (Figure 4). In contrast, CP29 contains three carotenoids: one Lut at the L1 site, one Neo at the N1 site, and one Vio at the L2 site, which is occupied by the second Lut in major LHCII. The V1 site in CP29 remains unoccupied[19,24]. CP29 has a long domain on its N-terminus not present in LHCII, and while major LHCII trimerizes in vivo, CP29 remains monomeric within the PSII supercomplex.

# Methods

## Sample Preparation

Reconstituted Lhcb1 and CP29 proteins were prepared according to Giuffra et al., Booth & Paulsen, Yang et al., and Natali et al.[1–4], with the following modifications. Briefly, *E. coli* BL21(DE3) pLysS cells harboring a pET-28a vector (Novagen) containing either the *LHCB1.3* gene from *Arabidopsis thaliana* or the *LHCB4* gene from *Spinacia oleracea* at the NcoI/XhoI sites were used to produce proteins with a C-terminal hexahistidine tag. The cells were inoculated into liquid LB medium containing 50 μg/mL kanamycin and grown overnight at 37°C with shaking. The resulting starter culture was used to inoculate 250 mL of Overnight Express liquid medium (MilliporeSigma) supplemented with 1% (v/v) glycerol and 50 μg/mL kanamycin. The culture was grown overnight at 37°C with shaking. Cells were harvested by centrifugation at 3,200 × *g* for 10 min at 23°C. The pellets were frozen at −20°C for 30 min and then thawed at room temperature. Inclusion bodies were isolated according to the protocol provided with the BugBuster Protein Extraction Reagent (MilliporeSigma). The total protein amount of the isolated inclusion bodies, resuspended in TE buffer [30 mM Tris-HCl (pH 8.0) and 1 mM EDTA], was quantified using the Pierce BCA Protein Assay Kit (Thermo Scientific).

Inclusion bodies containing 500 μg of CP29 or Lhcb1 apoprotein were solubilized in 100 mM HEPES-NaOH (pH 8.0), 75 mM sucrose, 2% (w/v) lithium dodecyl sulfate, and 10 mM DTT. The resuspended inclusion bodies were briefly vortexed, incubated at room temperature for 3 min, heated at 100°C for 1 min, and cooled to room temperature for 3 min. Reconstitution was carried out by combining 500 μg of apoprotein with 325 μg of pigment (275 μg of spinach whole-pigment extract and an additional 50 μg of spinach carotenoids at 10.8 μg/μL in 100% ethanol) in the presence of 1% (w/v) *n*-dodecyl-α-D-maltopyranoside (α-DM; Anatrace). The mixture was vortexed for 10–20 s at room temperature. After incubation on ice in the dark for 10 min, KCl was added to a final concentration of 150 mM, and the sample was vortexed for 30 s. Following another 10-min incubation on ice, the reconstituted sample was centrifuged at 15,800 × *g* for 10 min at 4°C.

The supernatant was collected and subjected to Ni-NTA affinity chromatography. Ni-NTA Superflow resin (Qiagen) was equilibrated with buffer containing 20 mM Tris-HCl (pH 8.0), 250 mM NaCl, 10% (w/v) glycerol, and 0.03% (w/v) α-DM. After His-tagged CP29 or Lhcb1 proteins were bound to the resin, the column was washed with 10 resin volumes of the same buffer supplemented with 15 mM imidazole. His-tagged proteins were eluted with five resin volumes of the same buffer supplemented with 400 mM imidazole. The eluted CP29 or Lhcb1 was further purified by gel filtration chromatography using a Superose 6 Increase 10/300 GL column connected to an ÄKTAmicro chromatography system (Cytiva), and the monomeric fraction was collected.

Buffer exchange into deuterated buffer [25 mM HEPES-NaOH (pH 7.8), 0.12 M NaCl, 0.3 M sucrose, and 0.03% (w/v) α-DM prepared in $D_2O$ (Sigma-Aldrich)] was performed by diluting the sample more than 100-fold in the deuterated buffer and then concentrating it using Amicon Ultra centrifugal concentrators with a 10-kDa molecular-weight cutoff. Pigment concentration was assessed by measuring the absorbance at 675 nm. Absorbance spectra from 350 to 750 nm were also recorded using an Infinite 200 PRO plate reader (Tecan) to assess sample quality.

**2DEV Measurements**

The 2DEV experimental setup has been previously described in detail[5–7]. The 2DEV spectra for both the CP29 and monomeric LHCII samples were taken with excitation pulses centered at 650 or 655 nm with approximately 70 nm FWHM (Figure S6). The repetition rate of the source laser was 1 kHz. Samples were diluted with d-glycerol to reach OD values of 0.7 at the 675 nm Chl *a* peak. Samples were then cooled to 85K using an Oxford Instruments Optistat cryostat to form a glass and minimize photobleaching. For both measurements, the path length was 250 μm, and excitation pulses were compressed to a FWHM of 18 fs. Both pulses were focused to a spot size of 200 μm with a combined energy of 155 nJ for CP29 and 190 nJ for monomeric LHCII samples.

The detection pulses were centered at 6100 nm and spanned a ~765 nm range. The instrument response function (IRF) is approximately 100 fs, estimated from the cross-correlation function between the mid-IR and visible pulses using a Germanium plate. The time delays between the second visible pump and the IR probe for CP29 are -50 fs to 1 ps in 25 fs steps, 1.5 to 10 ps in 0.5 ps steps, and 15 to 100 ps in 5 ps steps. The time delays for monomeric LHCII are -50 fs to 1 ps in 50 fs steps, 1.5 to 5 ps in 0.5 ps steps, 6 to 20 ps in 1 ps steps, and 25 to 30 ps in 5 ps steps.

**Structure-Based Kinetic Modeling**

The kinetic model was constructed following previous descriptions by Bennett et al., Leonardo et al., and Yang et al.[8–10] The models for both CP29 and monomeric LHCII are based on the 5XNL cryo-EM structure of the $C_2S_2M_2$-type PSII and subunit Hamiltonians constructed by Novoderezhkin and coworkers[11,12]. The zero-phonon line (ZPL) for each exciton state calculated in the literature needed to be shifted to the state energy from the diagonalized Hamiltonian. The average ZPL shift for CP29 was 255 cm$^{-1}$ for Chl *a* and 489 cm$^{-1}$ for Chl *b*. The average ZPL shift for Lhcb1 was 333 cm$^{-1}$ for Chl *a* and 469 cm$^{-1}$ for Chl *b*. We set the minimum exciton delocalization between sites to 0.15 for classification in the same domain, to prevent aphysical exciton delocalization in CP29 observed for a 0.1 cutoff. The simulation parameters are the same as in Leonardo et al.[8] and are listed in Table S1.

Spectral density definitions

The spectral density used for the simulation takes the form:

$$\chi''(\omega) = 2\lambda_0 \frac{\omega\Gamma_0}{\omega^2 + \Gamma_0^2}$$

Where $\lambda_0$ and $\gamma_0$ values for CP29 and Monomeric Lhcb1 are listed in table S_. An underdamped Brownian oscillator model is also used for vibronic coupling with individual modes, taking the following form:

$$\chi''_{vib}(\omega) = \sum_{j=1}^{N_{vib}} 2S_j\omega_j^3 \frac{\omega\Gamma_{vib}}{(\omega_j^2 - \omega^2)^2 + \omega^2\Gamma_{vib}^2}$$

$S_j$, $\omega_j$, and $\Gamma_j$ are parameters for each vibrational mode and can be found in refs[9,11].

**Table S1**. Kinetic model parameters including transition dipole moment (TDM), inhomogeneous broadening ($\sigma_{inhom}$, in $cm^{-1}$), and spectral density type (SD type).

| | | TDM | $\sigma_{inhom}$ | SD type |
|---|---|---|---|---|
| **CP29** | Chl *a* | 3.74 | 90 | a |
| | Chl *b* | 3.18 | 108 | b |
| **Lhcb1 Monomer** | Chl *a* | 4 | 80 | c |
| | Chl *b* | 3.4 | 96 | d |

**Table S2**. Spectral density parameters in units of $cm^{-1}$.

| | a | b | c | d |
|---|---|---|---|---|
| $\lambda_0$ | 40 | 48 | 37 | 48 |
| $\gamma_0$ | 40 | 40 | 30 | 30 |

The kinetic model was constructed at 85K to match the experimental condition. Based on the level of exciton delocalization and electronic coupling between states, we defined domains and applied modified Redfield theory within domains and generalized Forster theory between domains. Based on the inhomogeneous broadening widths reported in the literature (Table S1), a random number is added to the site energy of each chlorophyll, and 1000 site energies are generated to construct 1000 rate matrices.

The final rate matrix is the average of these inhomogeneous realizations. The use of the averaged rate matrix is supported by Amarnath et al.'s finding that the same fluorescence decay dynamics can be produced from the averaged rate matrix and from the individual rate matrices that are then averaged[13]. The simulation parameters are the same as in Leonardo et al.[8] and are listed in Table S2. To more accurately capture all energy transfer pathways present in the experiment, we add decay processes to the rate matrix. We set the average radiative rate across all excitons to 16 $ns^{-1}$ [14,15] and scale individual radiative rates for each exciton by their relative S0 – S1 transition dipole strengths. We also set the nonradiative rate for all excitons to 2 $ns^{-1}$ [14,16].

For each excitation frequency measured in the experiment, we predicted the initial population distribution across our simulated states based on the state energy and linewidth of each state. To most accurately replicate experimental conditions, we modeled the initial population arriving as a Gaussian pulse centered at 0 fs with $\sigma = 100$ fs. At each time step, we evolved the rate matrix using $P(t) = P(0)e^{Kt}$, where K is the exciton-to-exciton Markovian rate matrix in units of $ps^{-1}$.

**Lifetime Density Analysis (LDA)**

LDA uses a semi-continuous distribution of lifetimes to approximate a given data set:

$$\Delta A(t,\lambda) = \sum_{j=1}^{200} x_j(\tau_j,\lambda)e^{-t/\tau_j}$$

To minimize the effects of the ill-conditioned inverse Laplace transform while maintaining the ability to represent the data with many time constants, a semi-continuous distribution is employed in LDA. Regularization further minimizes computational instability and prevents overfitting. Here, we employ Tikhonov regularization[17], which penalizes large fitting coefficients, and takes the form:

$$min_{\vec{x}} \; || D\vec{x}_\lambda - \vec{A}_\lambda||_2^2 + \alpha||L\vec{x}_\lambda||_2^2$$

where L is the identity matrix, and $\alpha$ is the hyperparameter. $\alpha$ is important to select properly to find the optimal tradeoff between minimizing the residuals and the regression coefficients. In this work, we used $\alpha = 0.5$, which was determined by varying $\alpha$, performing LDA, reconstructing the signal time trace, and computing the $R^2$ value of the reconstructed signal compared to the original data (Figures S11 and S12).

In this work, we use 200 log-spaced time constants between 10 fs and 10 ps to perform LDA on both the experimental and simulated data sets. Past 10 ps, the signals reconstructed from LDA diverged from the experimental or simulated time traces (further details in the SI). We did not interpret LDA results for timescales less than 100 fs due to the experimental and simulated IRF of ~100 fs.

It is important to note that LDA is effective when extracting characteristic timescales of dynamics, but alone is not a reliable way to compute specific lifetime values for energy transfer processes. More accurate timescales may be obtained through fitting with global analysis, using the number of time components revealed by LDA[18]. For this work, we focus on the characteristic timescales because we are more interested in gaining a global understanding of patterns and behaviors in the energy transfer dynamics, rather than computing exact timescales of energy transfer between states.

In this work, we used $\alpha = 0.5$, which was determined by varying $\alpha$, performing LDA, reconstructing the signal time trace, and computing the $R^2$ value of the reconstructed signal compared to the original data (Figure S5). The $R^2$ values were largely the same for Lhcb1 across the $\alpha$ range of 0.2 – 1 and became worse with alpha values beyond these limits, but changed more significantly for CP29 across the same range. Although the $R^2$ values were higher for $\alpha = 0.2$ than for $\alpha = 0.5$ in the CP29 traces, it is clear how artifact can start to arise when alpha gets too low. In the time traces for both CP29 peaks, there is a general decay after ~0.5 ps for $\alpha \geq 0.5$. However, for the $\alpha = 0.2$ case, we observe an additional rise component between ~2-5 ps that is not present for larger alpha values. This indicates that although the $R^2$ value is higher for $\alpha = 0.2$ in this case, it is not the best choice for $\alpha$. Therefore, we use the $R^2$ values to find the general range of alpha for which the signal is best reconstructed, then choose the largest alpha that maintains accurate signal reconstruction. Ultimately, we chose $\alpha = 0.5$ to prevent overfitting the data while maintaining relatively good agreement with the raw data, but this value will depend on individual data sets and samples[25].

This deviation likely arises because inter-pigment energy transfer is largely complete by this time, leaving the signal dominated by slower population relaxation. As the magnitude of the remaining population dynamics decreases, the long-time signal contains less information with which to constrain the lifetime distribution, making the LDA solution increasingly sensitive to numerical error and noise. Consequently, small variations in the inferred lifetime distribution can produce larger deviations in the reconstructed dynamics.

Other factors may also contribute to the instability observed at longer times. As the signal amplitude decreases, the lifetime distribution becomes less well constrained, making the LDA solution more sensitive to noise and numerical error. In addition, long-lived exponential components can become increasingly difficult to distinguish from one another over the available time window, resulting in an ill-conditioned inverse problem. Numerical constraints or regularization imposed during the LDA may further influence the solution when the experimental signal provides limited information. Thus, the observed deviation after 10 ps may reflect a combination of reduced signal information, ill-conditioning, and increased sensitivity to numerical effects at long times. The sensitivity of LDA to perturbations is known[25].

SI References